\documentclass{svmult}
\usepackage{epsfig,graphicx} 
\usepackage[bottom]{footmisc}
\usepackage{natbib,url}      
\bibpunct{(}{)}{;}{a}{}{,}   
\def\cite#1{\citealp{#1}}    
\def\authorindex#1{}         
\def\figspath{.}

\begin{document}\newcount\preprintheader\preprintheader=1
{

\def\kms{km\,s$^{-1}$}


\def\thisvolume{these proceedings}

\def\aj{{AJ}}
\def\araa{{ARA\&A}}
\def\apj{{ApJ}}
\def\apjl{{ApJ}}
\def\apjs{{ApJS}}
\def\ao{{Appl.\ Optics}}
\def\apss{{Ap\&SS}}
\def\aap{{A\&A}}
\def\aapr{{A\&A~Rev.}}
\def\aaps{{A\&AS}}
\def\an{{Astron.\ Nachrichten}}
\def\aspcs{{ASP Conf.\ Ser.}}
\def\assp{{Astrophys.\ \& Space Sci.\ Procs., Springer, Heidelberg}}
\def\azh{{AZh}}
\def\baas{{BAAS}}
\def\jrasc{{JRASC}}
\def\memras{{MmRAS}}
\def\mnras{{MNRAS}}
\def\nat{{Nat}}
\def\pra{{Phys.\ Rev.\ A}}
\def\prb{{Phys.\ Rev.\ B}}
\def\prc{{Phys.\ Rev.\ C}}
\def\prd{{Phys.\ Rev.\ D}}
\def\prl{{Phys.\ Rev.\ Lett.}} 
\def\pasp{{PASP}}
\def\pasj{{PASJ}}
\def\qjras{{QJRAS}}
\def\science{{Sci}}
\def\skytel{{S\&T}}
\def\solphys{{Solar\ Phys.}}
\def\sovast{{Soviet\ Ast.}}
\def\ssr{{Space\ Sci.\ Rev.}}
\def\svassp{{Astrophys.\ Space Sci.\ Procs., Springer, Heidelberg}}
\def\zap{{ZAp}}
\let\astap=\aap
\let\apjlett=\apjl
\let\apjsupp=\apjs
\def\grl{{Geophys.\ Res.\ Lett.}}  
\def\jgr{{J. Geophys.\ Res.}} 

\def\ion#1#2{{\rm #1}\,{\uppercase{#2}}}  
\def\deg{\hbox{$^\circ$}}
\def\sun{\hbox{$\odot$}}
\def\earth{\hbox{$\oplus$}}
\def\la{\mathrel{\hbox{\rlap{\hbox{\lower4pt\hbox{$\sim$}}}\hbox{$<$}}}}
\def\ga{\mathrel{\hbox{\rlap{\hbox{\lower4pt\hbox{$\sim$}}}\hbox{$>$}}}}
\def\sq{\hbox{\rlap{$\sqcap$}$\sqcup$}}
\def\arcmin{\hbox{$^\prime$}}
\def\arcsec{\hbox{$^{\prime\prime}$}}
\def\fd{\hbox{$.\!\!^{\rm d}$}}
\def\fh{\hbox{$.\!\!^{\rm h}$}}
\def\fm{\hbox{$.\!\!^{\rm m}$}}
\def\fs{\hbox{$.\!\!^{\rm s}$}}
\def\fdg{\hbox{$.\!\!^\circ$}}
\def\farcm{\hbox{$.\mkern-4mu^\prime$}}
\def\farcs{\hbox{$.\!\!^{\prime\prime}$}}
\def\fp{\hbox{$.\!\!^{\scriptscriptstyle\rm p}$}}
\def\micron{\hbox{$\mu$m}}
\def\onehalf{\hbox{$\,^1\!/_2$}}
\def\onethird{\hbox{$\,^1\!/_3$}}
\def\twothirds{\hbox{$\,^2\!/_3$}}
\def\onequarter{\hbox{$\,^1\!/_4$}}
\def\threequarters{\hbox{$\,^3\!/_4$}}
\def\ubv{\hbox{$U\!BV$}}
\def\ubvr{\hbox{$U\!BV\!R$}}
\def\ubvri{\hbox{$U\!BV\!RI$}}
\def\ubvrij{\hbox{$U\!BV\!RI\!J$}}
\def\ubvrijh{\hbox{$U\!BV\!RI\!J\!H$}}
\def\ubvrijhk{\hbox{$U\!BV\!RI\!J\!H\!K$}}
\def\ub{\hbox{$U\!-\!B$}}
\def\bv{\hbox{$B\!-\!V$}}
\def\vr{\hbox{$V\!-\!R$}}
\def\ur{\hbox{$U\!-\!R$}}


\def\labelitemi{{\bf --}}

\def\rmit#1{{\it #1}}              
\def\rmit#1{{\rm #1}}              
\def\etal{\rmit{et al.}}           
\def\etc{\rmit{etc.}}
\def\ie{\rmit{i.e.,}}              
\def\eg{\rmit{e.g.,}}              
\def\cf{cf.}                       
\def\viz{\rmit{viz.}}
\def\vs{\rmit{vs.}}

\def\rot{\hbox{\rm rot}}
\def\div{\hbox{\rm div}}
\def\lesssim{\mathrel{\hbox{\rlap{\hbox{\lower4pt\hbox{$\sim$}}}\hbox{$<$}}}}
\def\gtrsim{\mathrel{\hbox{\rlap{\hbox{\lower4pt\hbox{$\sim$}}}\hbox{$>$}}}}
\def\mathstacksym#1#2#3#4#5{\def#1{\mathrel{\hbox to 0pt{\lower
    #5\hbox{#3}\hss} \raise #4\hbox{#2}}}}
\mathstacksym\lesssim{$<$}{$\sim$}{1.5pt}{3.5pt} 
\mathstacksym\gtrsim{$>$}{$\sim$}{1.5pt}{3.5pt} 
\mathstacksym\lrarrow{$\leftarrow$}{$\rightarrow$}{2pt}{1pt} 
\mathstacksym\lessgreat{$>$}{$<$}{3pt}{3pt} 

\def\dif{\: {\rm d}}                       
\def\ep{\:{\rm e}^}                        
\def\dash{\hbox{$\,-\,$}}                  
\def\is{\!=\!}                             

\def\starname#1#2{${#1}$\,{\rm {#2}}}  
\def\Teff{\hbox{$T_{\rm eff}$}}

\def\kms{\hbox{km$\;$s$^{-1}$}}
\def\ms{\hbox{m$\;$s$^{-1}$}}
\def\Mxcm{\hbox{Mx\,cm$^{-2}$}}    

\def\Bapp{\hbox{$B_{\rm app}$}}    

\def\komega{($k, \omega$)}                 
\def\kf{($k_h,f$)}                         
\def\VminI{\hbox{$V\!\!-\!\!I$}}           
\def\IminI{\hbox{$I\!\!-\!\!I$}}           
\def\VminV{\hbox{$V\!\!-\!\!V$}}           
\def\Xt{\hbox{$X\!\!-\!t$}}                

\def\level #1 #2#3#4{$#1 \: ^{#2} \mbox{#3} ^{#4}$}

\def\specchar#1{\uppercase{#1}}    
\def\AlI{\mbox{Al\,\specchar{i}}}  
\def\BI{\mbox{B\,\specchar{i}}}
\def\BII{\mbox{B\,\specchar{ii}}}
\def\BaI{\mbox{Ba\,\specchar{i}}}
\def\BaII{\mbox{Ba\,\specchar{ii}}}
\def\CI{\mbox{C\,\specchar{i}}}
\def\CII{\mbox{C\,\specchar{ii}}}
\def\CIII{\mbox{C\,\specchar{iii}}}
\def\CIV{\mbox{C\,\specchar{iv}}}
\def\CaI{\mbox{Ca\,\specchar{i}}}
\def\CaII{\mbox{Ca\,\specchar{ii}}}
\def\CaIII{\mbox{Ca\,\specchar{iii}}}
\def\CoI{\mbox{Co\,\specchar{i}}}
\def\CrI{\mbox{Cr\,\specchar{i}}}
\def\CriI{\mbox{Cr\,\specchar{ii}}}
\def\CsI{\mbox{Cs\,\specchar{i}}}
\def\CsII{\mbox{Cs\,\specchar{ii}}}
\def\CuI{\mbox{Cu\,\specchar{i}}}
\def\FeI{\mbox{Fe\,\specchar{i}}}
\def\FeII{\mbox{Fe\,\specchar{ii}}}
\def\FeIX{\mbox{Fe\,\specchar{ix}}}
\def\FeX{\mbox{Fe\,\specchar{x}}}
\def\FeXVI{\mbox{Fe\,\specchar{xvi}}}
\def\FrI{\mbox{Fr\,\specchar{i}}}
\def\HI{\mbox{H\,\specchar{i}}}
\def\HII{\mbox{H\,\specchar{ii}}}
\def\Hmin{\hbox{\rmH$^{^{_{\scriptstyle -}}}$}}      
\def\Hemin{\hbox{{\rm He}$^{^{_{\scriptstyle -}}}$}} 
\def\HeI{\mbox{He\,\specchar{i}}}
\def\HeII{\mbox{He\,\specchar{ii}}}
\def\HeIII{\mbox{He\,\specchar{iii}}}
\def\KI{\mbox{K\,\specchar{i}}}
\def\KII{\mbox{K\,\specchar{ii}}}
\def\KIII{\mbox{K\,\specchar{iii}}}
\def\LiI{\mbox{Li\,\specchar{i}}}
\def\LiII{\mbox{Li\,\specchar{ii}}}
\def\LiIII{\mbox{Li\,\specchar{iii}}}
\def\MgI{\mbox{Mg\,\specchar{i}}}
\def\MgII{\mbox{Mg\,\specchar{ii}}}
\def\MgIII{\mbox{Mg\,\specchar{iii}}}
\def\MnI{\mbox{Mn\,\specchar{i}}}
\def\NI{\mbox{N\,\specchar{i}}}
\def\NIV{\mbox{N\,\specchar{iv}}}
\def\NaI{\mbox{Na\,\specchar{i}}}
\def\NaII{\mbox{Na\,\specchar{ii}}}
\def\NaIII{\mbox{Na\,\specchar{iii}}}
\def\NeVIII{\mbox{Ne\,\specchar{viii}}}
\def\NiI{\mbox{Ni\,\specchar{i}}}
\def\NiII{\mbox{Ni\,\specchar{ii}}}
\def\NiIII{\mbox{Ni\,\specchar{iii}}}
\def\OI{\mbox{O\,\specchar{i}}}
\def\OVI{\mbox{O\,\specchar{vi}}}
\def\RbI{\mbox{Rb\,\specchar{i}}}
\def\SII{\mbox{S\,\specchar{ii}}}
\def\SiI{\mbox{Si\,\specchar{i}}}
\def\SiII{\mbox{Si\,\specchar{ii}}}
\def\SrI{\mbox{Sr\,\specchar{i}}}
\def\SrII{\mbox{Sr\,\specchar{ii}}}
\def\TiI{\mbox{Ti\,\specchar{i}}}
\def\TiII{\mbox{Ti\,\specchar{ii}}}
\def\TiIII{\mbox{Ti\,\specchar{iii}}}
\def\TiIV{\mbox{Ti\,\specchar{iv}}}
\def\VI{\mbox{V\,\specchar{i}}}
\def\HtwoO{\mbox{H$_2$O}}        
\def\Otwo{\mbox{O$_2$}}          

\def\Halpha{\mbox{H\hspace{0.1ex}$\alpha$}} 
\def\Ha{\mbox{H\hspace{0.2ex}$\alpha$}}
\def\Hbeta{\mbox{H\hspace{0.2ex}$\beta$}}
\def\Hgamma{\mbox{H\hspace{0.2ex}$\gamma$}}
\def\Hdelta{\mbox{H\hspace{0.2ex}$\delta$}}
\def\Hepsilon{\mbox{H\hspace{0.2ex}$\epsilon$}}
\def\Hzeta{\mbox{H\hspace{0.2ex}$\zeta$}}
\def\Lyalpha{\mbox{Ly$\hspace{0.2ex}\alpha$}}
\def\Lybeta{\mbox{Ly$\hspace{0.2ex}\beta$}}
\def\Lygamma{\mbox{Ly$\hspace{0.2ex}\gamma$}}
\def\Lycont{\mbox{Ly\hspace{0.2ex}{\small cont}}}
\def\Baalpha{\mbox{Ba$\hspace{0.2ex}\alpha$}}
\def\Babeta{\mbox{Ba$\hspace{0.2ex}\beta$}}
\def\Bacont{\mbox{Ba\hspace{0.2ex}{\small cont}}}
\def\Paalpha{\mbox{Pa$\hspace{0.2ex}\alpha$}}
\def\Bralpha{\mbox{Br$\hspace{0.2ex}\alpha$}}

\def\NaD{\mbox{Na\,\specchar{i}\,D}}    
\def\NaDone{\mbox{Na\,\specchar{i}\,\,D$_1$}}
\def\NaDtwo{\mbox{Na\,\specchar{i}\,\,D$_2$}}
\def\NaID{\mbox{Na\,\specchar{i}\,\,D}}
\def\NaIDone{\mbox{Na\,\specchar{i}\,\,D$_1$}}
\def\NaIDtwo{\mbox{Na\,\specchar{i}\,\,D$_2$}}
\def\Done{\mbox{D$_1$}}
\def\Dtwo{\mbox{D$_2$}}

\def\Mgbone{\mbox{Mg\,\specchar{i}\,b$_1$}}
\def\Mgbtwo{\mbox{Mg\,\specchar{i}\,b$_2$}}
\def\Mgbthree{\mbox{Mg\,\specchar{i}\,b$_3$}}
\def\MgIb{\mbox{Mg\,\specchar{i}\,b}}
\def\MgIbone{\mbox{Mg\,\specchar{i}\,b$_1$}}
\def\MgIbtwo{\mbox{Mg\,\specchar{i}\,b$_2$}}
\def\MgIbthree{\mbox{Mg\,\specchar{i}\,b$_3$}}

\def\CaIIK{\mbox{Ca\,\specchar{ii}\,K}}       
\def\CaIIH{\mbox{Ca\,\specchar{ii}\,H}}
\def\CaIIHK{\mbox{Ca\,\specchar{ii}\,H\,\&\,K}}
\def\HK{\mbox{H\,\&\,K}}
\def\Kthree{\mbox{K$_3$}}      
\def\Hthree{\mbox{H$_3$}}
\def\Ktwo{\mbox{K$_2$}}
\def\Htwo{\mbox{H$_2$}}
\def\Kone{\mbox{K$_1$}}
\def\Hone{\mbox{H$_1$}}
\def\KtwoV{\mbox{K$_{2V}$}}
\def\KtwoR{\mbox{K$_{2R}$}}
\def\KoneV{\mbox{K$_{1V}$}}
\def\KoneR{\mbox{K$_{1R}$}}
\def\HtwoV{\mbox{H$_{2V}$}}
\def\HtwoR{\mbox{H$_{2R}$}}
\def\HoneV{\mbox{H$_{1V}$}}
\def\HoneR{\mbox{H$_{1R}$}}

\def\hk{\mbox{h\,\&\,k}}
\def\kthree{\mbox{k$_3$}}
\def\hthree{\mbox{h$_3$}}
\def\ktwo{\mbox{k$_2$}}
\def\htwo{\mbox{h$_2$}}
\def\kone{\mbox{k$_1$}}
\def\hone{\mbox{h$_1$}}
\def\ktwoV{\mbox{k$_{2V}$}}
\def\ktwoR{\mbox{k$_{2R}$}}
\def\koneV{\mbox{k$_{1V}$}}
\def\koneR{\mbox{k$_{1R}$}}
\def\htwoV{\mbox{h$_{2V}$}}
\def\htwoR{\mbox{h$_{2R}$}}
\def\honeV{\mbox{h$_{1V}$}}
\def\honeR{\mbox{h$_{1R}$}}

\ifnum\preprintheader=1     
\makeatletter  
\def\@maketitle{\newpage
\markboth{}{}%
  {\mbox{} \vspace*{-8ex} \par
   \em \footnotesize To appear in ``Magnetic Coupling between the Interior
       and the Atmosphere of the Sun'', eds. S.~S.~Hasan and R.~J.~Rutten,
       Astrophysics and Space Science Proceedings, Springer-Verlag,
       Heidelberg, Berlin, 2009.} \vspace*{-5ex} \par
 \def\lastand{\ifnum\value{@inst}=2\relax
                 \unskip{} \andname\
              \else
                 \unskip \lastandname\
              \fi}%
 \def\and{\stepcounter{@auth}\relax
          \ifnum\value{@auth}=\value{@inst}%
             \lastand
          \else
             \unskip,
          \fi}%
  \raggedright
 {\Large \bfseries\boldmath
  \pretolerance=10000
  \let\\=\newline
  \raggedright
  \hyphenpenalty \@M
  \interlinepenalty \@M
  \if@numart
     \chap@hangfrom{}
  \else
     \chap@hangfrom{\thechapter\thechapterend\hskip\betweenumberspace}
  \fi
  \ignorespaces
  \@title \par}\vskip .8cm
\if!\@subtitle!\else {\large \bfseries\boldmath
  \vskip -.65cm
  \pretolerance=10000
  \@subtitle \par}\vskip .8cm\fi
 \setbox0=\vbox{\setcounter{@auth}{1}\def\and{\stepcounter{@auth}}%
 \def\thanks##1{}\@author}%
 \global\value{@inst}=\value{@auth}%
 \global\value{auco}=\value{@auth}%
 \setcounter{@auth}{1}%
{\lineskip .5em
\noindent\ignorespaces
\@author\vskip.35cm}
 {\small\institutename\par}
 \ifdim\pagetotal>157\p@
     \vskip 11\p@
 \else
     \@tempdima=168\p@\advance\@tempdima by-\pagetotal
     \vskip\@tempdima
 \fi
}
\makeatother     
\fi

\title*{Convection and the Origin of Evershed Flows}

\author{\AA. Nordlund\inst{1}
        \and
        G. B. Scharmer\inst{2}}
\authorindex{Nordlund, \AA.}
\authorindex{Scharmer, G. B.}
\institute{Niels Bohr Institute, University of Copenhagen, Denmark
           \and
           Institute for Solar Physics, Royal Swedish Academy of Sciences,
Stockholm, Sweden}

\maketitle
\setcounter{footnote}{0}  

\begin{abstract}
  Numerical simulations have by now revealed that the fine scale structure of
  the penumbra in general and the Evershed effect in particular is due to
  overturning convection, mainly confined to gaps with strongly reduced
  magnetic field strength. The Evershed flow is the radial component of the
  overturning convective flow visible at the surface.  It is directed
  outwards -- away from the umbra -- because of the broken symmetry due to the
  inclined magnetic field. The dark penumbral filament cores visible at high
  resolution are caused by the ``cusps'' in the magnetic field that form above
  the gaps.

  Still remaining to be established are the details of what determines
  the average luminosity of penumbrae, the widths, lengths, and filling
  factors of penumbral filaments, and the amplitudes and filling factors
  of the Evershed flow. These are likely to depend at least partially also
  on numerical aspects such as limited resolution and model size, but mainly
  on physical properties that have not yet been adequately determined or
  calibrated, such as the plasma beta profile inside sunspots at depth and
  its horizontal profile, the entropy of ascending flows in the penumbra,
  etc.
\end{abstract}

\section{Introduction}      \label{nordlund-sec:introduction}

Recently -- and in just the right time for the Evershed centenary -- the
first realistic three-dimensional radiation-magnetohydrodynamics models
of sunspots and sunspot penumbrae have become available
\citep{2007ApJ...669.1390H,2008ApJ...677L.149S,2009ApJ...691..640R}.  This has provided the
opportunity to resolve the longstanding debate about the origin of the
penumbral fine structure
\citep{%
1961ApJ...134..289D,%
1972SoPh...22..129M,%
1986A&A...158..351S,%
1993A&A...275..283S,%
1994A&A...290..295J,%
1996ApJ...463..372M,%
1998A&A...337..897S,%
1998ApJ...493L.121S,%
1999A&A...349..961S,%
2000A&A...361..734M,%
2002Natur.420..390T,%
2003A&A...403L..47B,%
2003A&A...411..257S,%
2004A&A...421..735S,%
2004A&A...422.1093B,%
2004A&A...427..319B,%
2004ApJ...600.1073W,%
2004ARA&A..42..517T,%
2004MNRAS.350..657T,%
2005A&A...436..333B,%
2006A&A...447..343S,%
2006A&A...452.1089T,%
2006A&A...460..605S,%
2008ApJ...686.1454B}
by ``looking the
horse in the mouth''.

When combined with unavoidable requirements from
basic physics, the evidence from even this first generation of numerical
models is sufficient to establish the basic mechanisms at work.  The models
illustrate, for example, that the large luminosity of the penumbra, of the
order of 75-95 \% of that of the photosphere \citep[][Fig.~3.2]{2003A&ARv..11..153S}, must be
essentially due to convective heat transport, which is able to carry nearly
as much heat to the solar surface in penumbrae as in the surrounding photosphere,
even in the presence of the strong penumbral magnetic field.

The numerical models still have shortcomings, as discussed in more detail
below.  In part, the shortcomings may be due to the choice of parameters
and boundary conditions for the models.  Additionally, the limited numerical
resolution of the models may also be important.  Indeed,
modelling entire sunspots is exceedingly demanding in computational resources
and even the most highly resolved models \citep[c.f.][]{2009ApJ...691..640R} have mesh
spacings that are only a few times smaller than the thickness of penumbral
filaments.

Since sunspots are not necessarily round one can also choose to model
only a narrow strip, stretching in one direction from an umbra to a
surrounding photosphere, and assumed to be periodic in the
perpendicular horizontal direction
\citep{2007ApJ...669.1390H,2009ApJ...691..640R}.  Such a model may be
considered to represent a piece of a sunspot with a section of more or
less straight penumbra filaments. There are abundant examples of such
sunspots, as well as of ``umbrae-without-penumbra'' and
``penumbrae-without-umbrae'', and thus one should expect to see
essentially the same phenomena in such models as in models of entire
sunspots.  The much reduced size requirement in one direction can be
utilized to increase the resolution in the other two directions, with
given amounts of computing resources.

As we show below the situation can also be modelled, with essential features
reproduced, in even smaller patches, with accordingly higher possible spatial
resolution, by making also the longer of the two horizontal directions
periodic.  Such models represent local rectangular patches of penumbrae, and
include neither an umbra nor a piece of photosphere.  The one major drawback of
this type of model is that it cannot allow for the inclination of the penumbra
surface.  The setup does allow for an arbitrary inclination of the penumbra
magnetic field, and in fact requires the inclination to be specified, at least
as an initial condition. These types of models are thus ideal for exploring the
parameter space spanned by magnetic field inclination and by lower boundary
plasma beta and entropy, and may be used to disentangle the relative importance
of each these factors, at spatial resolutions that could not be achieved in the
other two types of models.

The three classes of models are complementary, in that the more localized
ones can achieve higher spatial resolution, and thus can address resolution
issues more effectively than global models, which on the other hand can
be compared more directly with observations.

Below we briefly summarize and discuss some of the findings from the
numerical models, attempt to identify reasons for remaining problems,
and indicate directions for future work.

\section{Recent modelling results}      \label{nordlund-sec:recent}

\citet{2007ApJ...669.1390H} were the first to perform realistic numerical simulations
of sunspots where one could see signs of the formation of penumbra filaments.
The qualifier ``realistic'' is here used to indicate MHD-models that include
an ionizing equation of state representative of solar conditions and radiative
energy transfer with an H$^-$-like continuum opacity.  Those requirements are
essential for making sure that the model behavior near the optical depth
unity, where the surface luminosity is determined, is at least qualitatively
similar to that of the real Sun.  Gray radiative transfer was used in the
Pencil Code results on which the \citet{2007ApJ...669.1390H} paper was based, but
similar experiments using the Copenhagen Stagger Code that includes non-gray
radiative energy transfer
\begin{figure}
  \centering
  \centering\includegraphics[width=0.8\textwidth]{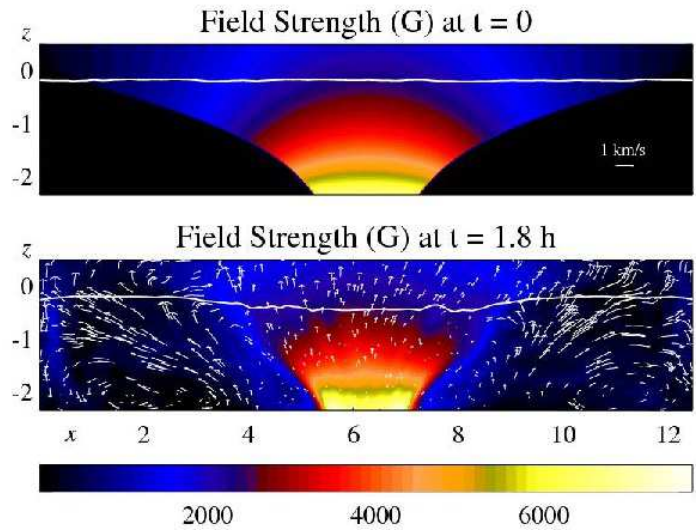}
  \caption[]{\label{nordlund-fig:1} Cross sections of the
  \citet{2007ApJ...669.1390H} experiment, showing the initial state (top), where
  the magnetic field has a potential shape corresponding to a constant beta
  as a function of height, and the state after 1.8 solar hours (bottom), where the
  gas pressure has been significantly reduced in the surface layers, because
  of the cooling of the surface layers associated with the strong magnetic field.
  The arrows shown in the bottom panel illustrate the ``moat flow''; a mean
  sub-surface radial flow driven by the reduced cooling under the spot.
}\end{figure}
\begin{figure}
  \centering\includegraphics[width=0.8\textwidth]{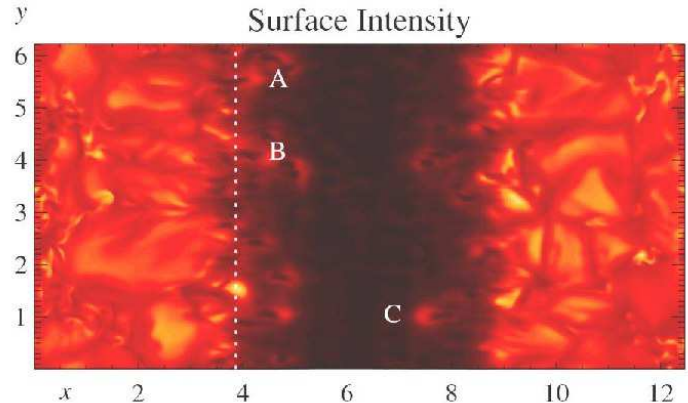}
  \caption[]{\label{nordlund-fig:2} Image of the surface radiation intensity
  from the \citet{2007ApJ...669.1390H} experiment, showing a number of
  short penumbral filaments with dark cores.  Three of the locations are
  marked, and are further illustrated in Fig.~\ref{nordlund-fig:3}.
  Note the small scale bright (magnetic) features outside the spot -- in
  animations they can be seen drifting out from the spot, carried by
  the moat flow shown in Fig.~\ref{nordlund-fig:1}.
}\end{figure}
confirm the results.  The use of non-gray (binned) opacity only influences layers
higher up in the atmosphere, whose behavior is not essential to understanding
the penumbra fine structure (although it may be important for computing
realistic synthetic spectral line diagnostics).

The \citet{2007ApJ...669.1390H} experiments consisted of a rectangular box of size
$12\times6\times3$ Mm, with an isotropic grid spacing of approximately 24 km; a
vertical cross section is shown in Fig.~\ref{nordlund-fig:1}. As users of
numerical MHD-simulations are the first to realize this is by no means
equivalent to having a ``physical resolution'' of 24~km; as a rule-of-thumb a
feature resolved by less than about 10 grid points is significantly affected by
numerical resolution, and hence the penumbral small scale features seen in the
\citet{2007ApJ...669.1390H} models are indeed only marginally resolved.  One can
nevertheless recognize a number of properties reminiscent of solar penumbrae
and their surroundings:
bright filaments with ``heads'' that move towards the umbra, with dark cores
(cf.\ Fig.~\ref{nordlund-fig:2}) and systematic radial outflows of
several {\kms} in the penumbral filaments, and a ``moat flow'' carrying magnetic
flux features away from the spot.  Both the speeds of inward migration, the
magnitude of the Evershed-like outflows, and the moat flow speeds are similar
to observed values.  Those agreements could in principle, given the limited
spatial size and resolution of the model, be fortuitous, but could also be
indications of underlying physical processes that are so robust that they start
becoming established even when the numerical resolution is marginal.  As
illustrated by the subsequent work by \citet{2009ApJ...691..640R} the latter indeed seems
to be the case.

\begin{figure}
  \centering\includegraphics[width=0.8\textwidth]{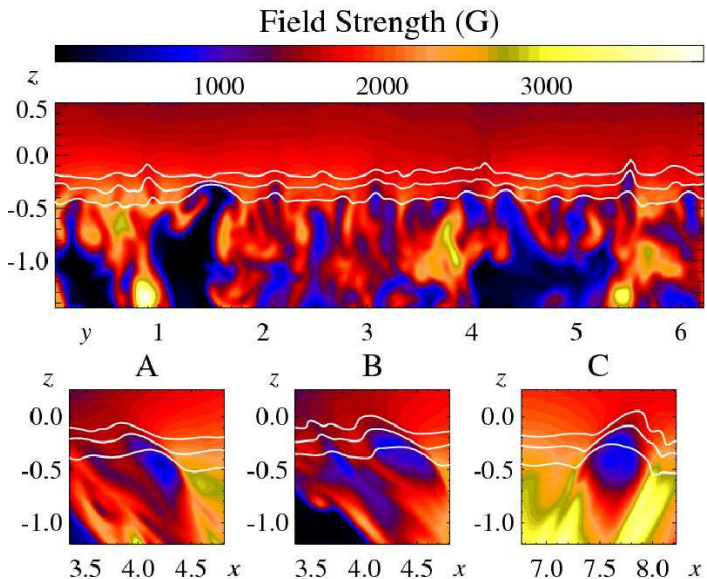}
  \caption[]{\label{nordlund-fig:3} Cross sections showing the magnetic
  field strength along the dotted line, and across the three features indicated
  in Fig.~\ref{nordlund-fig:2}. These are migrating penumbral features, with
  strong upflows on their umbral sides and downflows on their penumbral
  sides.
}\end{figure}

As pointed out by \citet{2007ApJ...669.1390H}, the filamentary features seen at
the surface correspond to local convective flow channels (cf.\
Fig.~\ref{nordlund-fig:3}), similar to the umbral dots seen in the
numerical simulations be \citet{2006ApJ...641L..73S}, and in light bridge
models \citep[][ cf.\ Fig. \ref{nordlund-fig:9}]{2006ASPC..354..353N}.
The analogy is additional
evidence of robust physical processes at work.  The main difference is probably
just a matter of the inclination of the magnetic field, which brakes the
symmetry and forces a preferred direction onto the overturning part of the
convective flow.

The basic mechanism at work is thermal convection, modified by the
presence of a strong and inclined magnetic field.  The luminosity of the
penumbra, which visually is so clearly attributed to the densely packed
penumbral filaments, must be due to convective heat transport up to the visible
surface, with the penumbral filaments being the top layer manifestation of that
heat transport, much as the granulation pattern in the surrounding photosphere
is the surface manifestation of the even more efficient convective heat
transport there.

\begin{figure}
  \centering\includegraphics[width=0.8\textwidth]{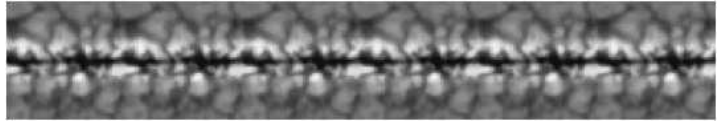}
  \caption[]{\label{nordlund-fig:9} Light bridge formed by a ``gap'' in an
  umbral-strength magnetic field.   As a consequence of approximate
  horizontal pressure equilibrium the gap is cusp-shaped, closing at the
  top at some height in the visible photosphere.  Because of the
  correspondingly higher density of the gas inside the cusp and the
  overall drop of temperature with height, a dark filament-like
  structure appears in the emergent intensity
  (cf.\ \cite{2006ASPC..354..353N}).
}\end{figure}

In the penumbra the overturning and cooling of ascending hot gas is constrained
to happen primarily in the radial direction; the generally radial orientation
of the magnetic field strongly disfavors motions in the azimuthal direction.
Since each parcel of ascending hot gas must be exposed to surface cooling for
about the same time as in the granulation pattern in the surrounding
photosphere, and since the velocity magnitudes are constrained by the
requirement to transport nearly the same amount of energy to the surface, the
trajectories of parcels of gas as they overturn at the surface must be about as
long in the surrounding photosphere; i.e., paths that extend above the surface
for several megameters horizontally.   But since the motions are constrained
to occur mainly in the radial direction there is less horizontal expansion
and more ``crowding'' of the paths.

From these general considerations one concludes that already the energy
transport requirements and the constraints imposed by the magnetic field
in a natural way leads to the occurrence of bright, densely packed filaments
with lengths of at least several megameters.  To the extent that the models
do not yet produce all of these properties (e.g., too short and
non-space-filling filaments) one can then also conclude that the energy
transport (and hence penumbral brightness) is likely to be too small,
for which one should then seek remedies.

However, the features that are already present in the current models do perform the expected
type of convective energy transport, just at a lower level of intensity, and
with a smaller filling factor.  One can thus already study the basic mechanisms
at work, and perhaps then also conjecture about how conditions should be
changed in order for the models to more close match the observed properties of
solar penumbrae.

\begin{figure}
  \centering\includegraphics[width=0.8\textwidth]{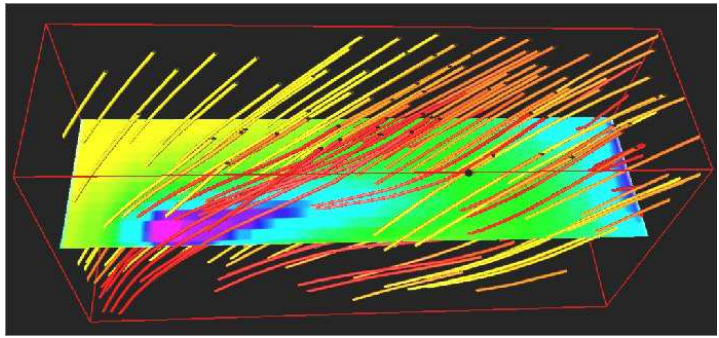}
  \caption[]{\label{nordlund-fig:4} This snapshot of a 3-D visualization
  \citep{2008ApJ...677L.149S} shows the vector magnetic field, color coded with
  magnetic field strength, with a cutting plane containing a color coded image
  of temperature. Purple to blue represents large values, orange to red
  represent small values (color illustrations are available in the on-line
  version). The size of the box is approx.\ $1200\times 340\times 550$ km.
}\end{figure}


\subsection{Evershed results}
\citet{2008ApJ...677L.149S} investigated the nature of the Evershed-like flows in
the penumbral filaments seen in the \citet{2007ApJ...669.1390H} models, looking
in particular for explanations of the inward migration and for the outward,
Evershed-like flows.  They found the inward migration to be due to a
pattern-motion where the ``head'' represents a strong convective upflow
location, which is able to push aside the penumbral magnetic field,
and also to push material that then quickly cools up along the inclined
magnetic field lines.  The cooler and heavier material results in the
bending down of field lines, and allows the convective flow to overturn
and return down below the surface.  That process, the ascent of hot gas
that cools at the surface and then descends, is -- from a ``heat engine'' point
of view -- essentially the same as the convective process in normal solar
granulation \citep[cf.\ ][]{1998ApJ...499..914S,2008PhST..133a4002N}.

The filaments seen in the \citet{2007ApJ...669.1390H} simulations are quite short,
which is perhaps not too surprising, given the limitations in size and
numerical resolution.  The corresponding features in the models of
\citet{2009ApJ...691..640R} are longer, but still rather short compared to real
penumbral filaments.   In both cases the filaments are also much less
space-filling than in real solar penumbrae.  Since the size
constraints have been lifted in the \citet{2009ApJ...691..640R}
work the length and filling factor issues should be seen as indications
that some other setup property common to both works is responsible;
e.g., the limited numerical resolution.
\citet{2009ApJ...691..640R} also argue that the Evershed-like flow seen in their
model is too weak relative to real sunspots, but this may simply be a
reflection of the too small filling factor.  As illustrate by Fig.~6
of their paper, the peak flow speeds in the filaments reach approx.\ 6 {\kms},
which is certainly in the right ball park \citep[cf.][]{2003A&ARv..11..153S}.

\begin{figure}
  \centering\includegraphics[width=0.8\textwidth]{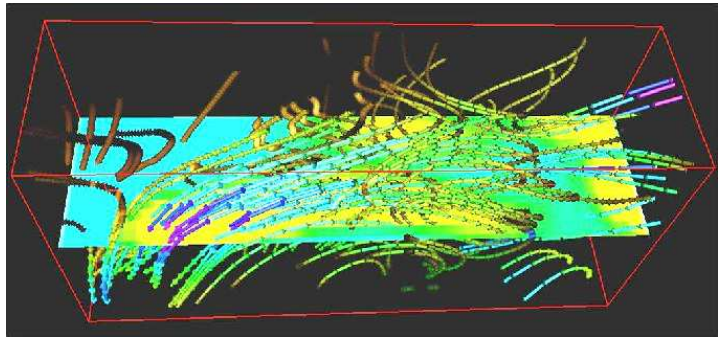}
  \caption[]{\label{nordlund-fig:6} The vector velocity field, color
    coded with velocity magnitude, and with a cutting plane
    color-coded image showing the strength of the vertical magnetic
    field component.
}\end{figure}

Likely factors, which could be responsible for both the lack of space
filling and the shortness of the filaments, are the basic strength and
inclinations of the magnetic field in the penumbral part of the
models. These properties depend on details of the boundary conditions
applied at the base of the model -- mainly the amplitude and profile
of the plasma beta (ratio of gas to magnetic pressure) across the base
of the sunspot model.  Already a visual inspection of
Fig.~\ref{nordlund-fig:1}, and of Fig.~2 of
\citet{2009ApJ...691..640R} indicates that the average model magnetic
fields are not sufficiently inclined, as compared to what is implied
by Fig.~3.2 of \citet{2003A&ARv..11..153S}.  Likely remedies that
should be tried include decreasing the sunspot beta value at the
bottom boundary, adjusting the horizontal beta profile, and/or
reducing the sensitivity to boundary conditions by extending the
models to larger depths.  Future model results will show if any of
these suggestions pans out, or if there are yet other properties of
the models that need to be improved (e.g., using even higher spatial
resolution, and continuing for even longer time spans, sufficient for
the model to forgive and forget oversimplified initial conditions).

\section{Local penumbra models}                \label{nordlund-sec:local}
As mentioned in the introduction, one of the compromises one can choose to
make to maximize the available spatial resolution is to restrict the total
size of the model so much that it represents only a small piece of the
penumbra.  To avoid having to specify necessarily awkward radial
boundary conditions one can choose periodic boundary conditions also
in the radial direction.

Such models necessarily lack one seemingly central aspect of real penumbrae, in
that the sloping of the penumbra surface (e.g., of optical depth unity) cannot
be represented.  That may also be seen as an advantage, since in the
alternative, more normal set-ups, which include an umbra and a photosphere,
one cannot avoid having a sloping penumbra.  By using local models one can
thus figure out if the sloping plane is essential for the formation of
the penumbral fine structure, or if perhaps -- as in the case of rectangular
vs.\ round sunspot models -- this is a side issue that is not of central
importance.

\begin{figure}
  \centering\includegraphics[width=0.8\textwidth]{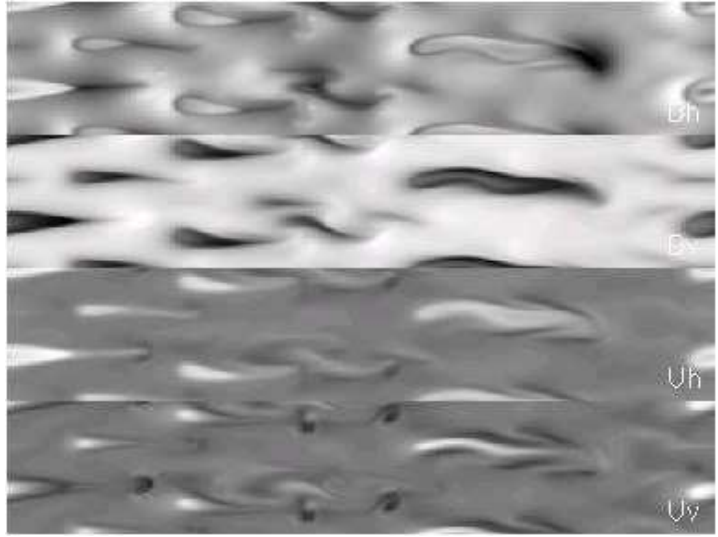}
  \caption[]{\label{nordlund-fig:7} A stack of horizontal cross sections
  near mean optical depth unity from a local penumbra model with 3 km grid
  spacing.  The panels show gray scale images of the horizontal and vertical
  magnetic field strengths and velocity amplitudes.
}\end{figure}

The main advantage with the local models is that they allow a much larger
spatial resolution.  A rectangular model of a penumbral stripe such as
the \citet{2007ApJ...669.1390H} model needs to be at least as wide as a few granules,
in order to do a reasonable job in representing the external photosphere.
At the same time it needs to be at least as ``long'' as the diameter of a sunspot,
and then some, to allow for the surrounding photosphere and to prevent
neighboring periodic images of the spot from distorting the spot magnetic field
too much.  The
\citet{2007ApJ...669.1390H} model, which was $12\times 6$ Mm wide, thus represents
the low end of the possible size range of these models.

A local and doubly periodic model that covers only a patch of the penumbra,
on the other hand needs only to be as wide as a few penumbral filaments
(say, about 1 Mm), and long enough to contain a section of a penumbral
filament (which was argued above to be of the order of the size of a
granule, so a size of about 6 Mm would seem to be enough).  To summarize
one can use a horizontal area more than an order of magnitude smaller in
local models, with a corresponding gain possible in the spatial resolution.
Figure \ref{nordlund-fig:7} shows an example of such a model, with a grid
spacing of about 3 km.

The filamentary structures visible in Figure \ref{nordlund-fig:7} display a
pattern motion towards the penumbra, consistent with the behavior found in
similar, but more idealized setups that have been investigated in the past
\citep[cf.][]{2000SoPh..192..109H}.  This demonstrates that pattern motions
similar to those observed in the inner parts of real penumbra can occur also
without a `sloping plane' penumbra surface.  Systematic studies of the effects
of varying the strength and inclination of the magnetic field in such models
will be helpful in disentangling the interplay between factors that influence
the lengths, widths, and filling factors of small scale penumbral structures.

From the various results mentioned above one can draw some conclusions relevant
to the debate about mechanisms (or mechanism labels!) responsible for the
creation of penumbral filaments.  A major label issue has for example
been whether to refer to the penumbral structures as ``flux tubes'' or
as ``gaps''.

To some extent such issues become moot when detailed models become available,
or at most they turn into ``what-did-I-say'' attempts to paste the labels onto
the detailed models.
It is, however, relevant to remark that the flows that occur in penumbral
filaments (and this includes the Evershed flow) are flows that -- at least
locally -- dominate over the magnetic fields. The overturning convective flows
are able to push aside the magnetic field, weakening it sufficiently so it no
longer prevents the flow, but allows it.  In this respect the situation is
more adequately labelled as characterized by magnetic ``gaps'', rather than
one characterized by magnetic ``flux tubes'', in as much as ``flux tubes''
traditionally have been though of as magnetic field channels that
constrain the motions of gas to be along the magnetic field in the structures,
rather than flowing relatively freely into and out of them, carrying a relatively
weak magnetic field along.  In the penumbra, the filaments do represent structures
that are ``open'' (at the bottom) rather than ``closed'',
and where the topology of the flow determines
the topology of the magnetic field rather than the other way around.  Of
course the gaps are not entirely ``field free'', but the they fulfil essentially
the same function as the idealized field-free gaps originally proposed by
\citet{2006A&A...447..343S} and \citet{2006A&A...460..605S}.

With respect to the ``siphon flow'' type of models \citep{%
1989ApJ...337..977M,%
1990ApJ...359..550T,%
1991ApJ...375..404T,%
1993ApJ...402..314M,%
1997Natur.390..485M,%
2005A&A...440L..29T}
it appears
safe to conclude that they are not directly relevant  for penumbral fine
structure.  It is possible that localized siphon flows do occur, e.g., along
field lines that happen to connect locations inside the penumbra with strongly
evacuated flux concentrations outside the sunspot, but these must be rare, and
cannot be a significant contributing effect in the creation and maintenance of
penumbral fine structure.

The idea that downward flux pumping \emph{outside} the sunspot could
contribute to creating and maintaining the penumbra fine structure, by
``pulling down'' field lines and hence helps explain the occurrence of nearly
horizontal field lines in the penumbra \citep{%
2002Natur.420..390T,%
2004ApJ...600.1073W}
also does not work, or at least is not needed.  However, \emph{inside} the penumbra,
pulling down of field lines by convection is certainly relevant and
operative, and near the periphery of penumbrae some aspect of such
a process may be relevant as well \citep[cf.][]{2008ApJ...686.1454B}.

An interesting question, which can be more easily addressed with the local type
of penumbra models that allow maximizing the numerical resolution, is whether
and to what extent the reduction of the magnetic field strength in the gaps depends on the
numerical resolution.  Should one expect, for example, that the magnetic field
strength inside the gaps is reduced to a larger and larger extent as the
numerical resolution is increased in the models?  Only future experiments can
address that question quantitatively, but two fundamental aspects deserve to be
mentioned already before the results of such investigations are in:
\begin{itemize}
\item To the extent that the flows are of finite duration, and their
source regions are not entirely field free (both assumptions are unavoidably
satisfied in the penumbral context) there is a definitive limit to how weak the
magnetic fields can get, in that even with ``frozen-in'' field lines the
magnetic flux density is only reduced by an amount corresponding to the
perpendicular expansion of the flow from the ``source'' point to the point of
observation.
\item Moreover, a general conclusion from other situations with very
low diffusivities is that some sort of turbulence typically develops,
resulting in large scale statistics that become essentially independent
of the scale at which diffusivity sets in.  To the extent that this happens here
one should not expect a continued dependence on the numerical resolution.
\end{itemize}

\section{Conclusions and concluding remarks}      \label{nordlund-sec:concl}
In conclusion, with access to realistic three-dimensional numerical modeling of
sunspots we are now in an excellent position to finally answer the intriguing
questions raised by Evershed's observations a hundred years ago
\citep{1909Obs....32..291E,1909MNRAS..69..454E}. Accurate and
detailed matches and comparisons with observations must await the next
generation of improved models, but the current set of models already allow a
number of conclusions regarding the physical mechanisms at work in the
penumbra:
\begin{itemize}
\item The basic mechanism, responsible for creating the penumbra fine structure
as well as the Evershed flow, is thermal convection, modified by the
presence of a strong and inclined magnetic field, with the penumbral filaments
being the top layer manifestation of the heat transport, analogous to the
convection pattern in the surrounding photosphere.
\item In the penumbra the overturning and cooling of ascending hot gas is
constrained to happen primarily in the radial direction, with the inclination
of the magnetic field braking the symmetry and causing the overturning to
happen as an outward directed mainly radial flow -- this is the physical
background for the Evershed flow.
\item The convective motions are able to force open ``gaps'' in the magnetic
field. These gaps are of course not ``field-free'', but the magnetic field
strengths are reduced enough to allow the overturning convective flows to
occur.
\item Energy transport requirements dictate that ascending hot parcels of gas
spend similar intervals of time, with similar flow speeds, while cooling in the
optically visible surface layers as they do in the photospheric granulation
pattern.  Individual sections of penumbral filaments must therefore have lengths
similar to granulation scales, and there both upflows and downflows must occur,
intertwined, in the penumbral filaments.
\end{itemize}

With qualitative and semi-quantitative agreement of the results of realistic
numerical simulations from two independent groups
\citep{2007ApJ...669.1390H,2009ApJ...691..640R}, and with quantitative results moving in the
right direction with improvements in the numerical resolution and in
the horizontal and vertical extent of the models, the results that have
already been gleaned from the first generation of models are likely to
hold up to future scrutiny.  But it will indeed be interesting to see models
improving to the point where direct, quantitative comparisons can be made
between observations and synthetic diagnostic produced from the models.

The intricate correlations between magnetic field strength and inclination
on the one hand and velocity, temperature, and gas density on the other
hand makes it virtually impossible to uniquely ``invert'' even very detailed
observations.  Indeed, the more detailed the observations are (in the spatial,
temporal, and spectral domains) the more parameters it would take to
represent the physical properties of inversion type models.

On the other hand, with access to detailed numerical models one can compute
synthetic spectral lines and related diagnostics such as synthetic magnetograms
and Doppler maps, and compare both their qualitative features and more
quantitative ``finger prints'', such as average spectral line shapes and other
statistical measures.  Each and every such quantitative measure represents a
``collapse'' of information, making it possible to compare unique signatures of
the complex reality with analogous results from (necessarily always less
complex) numerical models.
If accurate matches are obtained one may still argue that this does not constitute
formal proof of correctness of the models, but to the extent that the models
adhere closely to the physical properties (equation of state, realistic
radiative energy transfer, etc.) of the situation they are representing one
would be hard put to argue that a good match is inconclusive.

\begin{acknowledgement}
  We thank the conference organizers for a very good meeting and the editors
  for excellent instructions and an amazing patience.  Discussions and
  collaborations with Henk Spruit and Tobias Heinemann are gratefully
  acknowledged. The work of {\AA}N is supported by the Danish Natural Science
  Research Council. Computing resources were provided by the Danish Center for
  Scientific Computing.
\end{acknowledgement}

\begin{small}

\bibliographystyle{rr-assp}       

\begin{thebibliography}{40}
\expandafter\ifx\csname natexlab\endcsname\relax\def\natexlab#1{#1}\fi

\bibitem[{{Bellot Rubio} {et~al.}(2004){Bellot Rubio}, {Balthasar}, \&
  {Collados}}]{2004A&A...427..319B}
{Bellot Rubio}, L.~R., {Balthasar}, H., {Collados}, M. 2004, \aap, 427, 319

\bibitem[{{Bellot Rubio} {et~al.}(2003){Bellot Rubio}, {Balthasar}, {Collados},
  \& {Schlichenmaier}}]{2003A&A...403L..47B}
{Bellot Rubio}, L.~R., {Balthasar}, H., {Collados}, M., {Schlichenmaier}, R.
  2003, \aap, 403, L47

\bibitem[{{Borrero} {et~al.}(2005){Borrero}, {Lagg}, {Solanki}, \&
  {Collados}}]{2005A&A...436..333B}
{Borrero}, J.~M., {Lagg}, A., {Solanki}, S.~K., {Collados}, M. 2005, \aap, 436,
  333

\bibitem[{{Borrero} {et~al.}(2004){Borrero}, {Solanki}, {Bellot Rubio}, {Lagg},
  \& {Mathew}}]{2004A&A...422.1093B}
{Borrero}, J.~M., {Solanki}, S.~K., {Bellot Rubio}, L.~R., {Lagg}, A.,
  {Mathew}, S.~K. 2004, \aap, 422, 1093

\bibitem[{{Brummell} {et~al.}(2008){Brummell}, {Tobias}, {Thomas}, \&
  {Weiss}}]{2008ApJ...686.1454B}
{Brummell}, N.~H., {Tobias}, S.~M., {Thomas}, J.~H., {Weiss}, N.~O. 2008, \apj,
  686, 1454

\bibitem[{{Danielson}(1961)}]{1961ApJ...134..289D}
{Danielson}, R.~E. 1961, \apj, 134, 289

\bibitem[{{Evershed}(1909{\natexlab{a}})}]{1909Obs....32..291E}
{Evershed}, J. 1909{\natexlab{a}}, The Observatory, 32, 291

\bibitem[{{Evershed}(1909{\natexlab{b}})}]{1909MNRAS..69..454E}
{Evershed}, J. 1909{\natexlab{b}}, \mnras, 69, 454

\bibitem[{{Heinemann} {et~al.}(2007){Heinemann}, {Nordlund}, {Scharmer}, \&
  {Spruit}}]{2007ApJ...669.1390H}
{Heinemann}, T., {Nordlund}, {\AA}., {Scharmer}, G.~B., {Spruit}, H.~C. 2007,
  \apj, 669, 1390

\bibitem[{{Hurlburt} {et~al.}(2000){Hurlburt}, {Matthews}, \&
  {Rucklidge}}]{2000SoPh..192..109H}
{Hurlburt}, N.~E., {Matthews}, P.~C., {Rucklidge}, A.~M. 2000, \solphys, 192,
  109

\bibitem[{{Jahn} \& {Schmidt}(1994)}]{1994A&A...290..295J}
{Jahn}, K., {Schmidt}, H.~U. 1994, \aap, 290, 295

\bibitem[{{Mamadazimov}(1972)}]{1972SoPh...22..129M}
{Mamadazimov}, M. 1972, \solphys, 22, 129

\bibitem[{{Martens} {et~al.}(1996){Martens}, {Hurlburt}, {Title}, \&
  {Acton}}]{1996ApJ...463..372M}
{Martens}, P.~C.~H., {Hurlburt}, N.~E., {Title}, A.~M., {Acton}, L.~W. 1996,
  \apj, 463, 372

\bibitem[{{Mart{\'{\i}}nez Pillet}(2000)}]{2000A&A...361..734M}
{Mart{\'{\i}}nez Pillet}, V. 2000, \aap, 361, 734

\bibitem[{{Montesinos} \& {Thomas}(1989)}]{1989ApJ...337..977M}
{Montesinos}, B., {Thomas}, J.~H. 1989, \apj, 337, 977

\bibitem[{{Montesinos} \& {Thomas}(1993)}]{1993ApJ...402..314M}
{Montesinos}, B., {Thomas}, J.~H. 1993, \apj, 402, 314

\bibitem[{{Montesinos} \& {Thomas}(1997)}]{1997Natur.390..485M}
{Montesinos}, B., {Thomas}, J.~H. 1997, \nat, 390, 485

\bibitem[{{Nordlund}(2006)}]{2006ASPC..354..353N}
{Nordlund}, {\AA}. 2006, in Solar MHD Theory and Observations: A High Spatial
  Resolution Perspective, eds. J.~{Leibacher}, R.~F. {Stein}, \&
  H.~{Uitenbroek}, \aspcs, 354,
  353

\bibitem[{{Nordlund}(2008)}]{2008PhST..133a4002N}
{Nordlund}, {\AA}. 2008, Physica Scripta Vol.\ T, 133, 014002

\bibitem[{{Rempel} {et~al.}(2009){Rempel}, {Sch{\"u}ssler}, \&
  {Kn{\"o}lker}}]{2009ApJ...691..640R}
{Rempel}, M., {Sch{\"u}ssler}, M., {Kn{\"o}lker}, M. 2009, \apj, 691, 640

\bibitem[{{Scharmer} {et~al.}(2008){Scharmer}, {Nordlund}, \&
  {Heinemann}}]{2008ApJ...677L.149S}
{Scharmer}, G.~B., {Nordlund}, {\AA}., {Heinemann}, T. 2008, \apjl, 677, L149

\bibitem[{{Scharmer} \& {Spruit}(2006)}]{2006A&A...460..605S}
{Scharmer}, G.~B., {Spruit}, H.~C. 2006, \aap, 460, 605

\bibitem[{{Schlichenmaier} {et~al.}(1999){Schlichenmaier}, {Bruls}, \&
  {Sch{\"u}ssler}}]{1999A&A...349..961S}
{Schlichenmaier}, R., {Bruls}, J.~H.~M.~J., {Sch{\"u}ssler}, M. 1999, \aap,
  349, 961

\bibitem[{{Schlichenmaier} {et~al.}(1998{\natexlab{a}}){Schlichenmaier},
  {Jahn}, \& {Schmidt}}]{1998ApJ...493L.121S}
{Schlichenmaier}, R., {Jahn}, K., {Schmidt}, H.~U. 1998{\natexlab{a}}, \apjl,
  493, L121+

\bibitem[{{Schlichenmaier} {et~al.}(1998{\natexlab{b}}){Schlichenmaier},
  {Jahn}, \& {Schmidt}}]{1998A&A...337..897S}
{Schlichenmaier}, R., {Jahn}, K., {Schmidt}, H.~U. 1998{\natexlab{b}}, \aap,
  337, 897

\bibitem[{{Schlichenmaier} \& {Solanki}(2003)}]{2003A&A...411..257S}
{Schlichenmaier}, R., {Solanki}, S.~K. 2003, \aap, 411, 257

\bibitem[{{Schmidt} {et~al.}(1986){Schmidt}, {Spruit}, \&
  {Weiss}}]{1986A&A...158..351S}
{Schmidt}, H.~U., {Spruit}, H.~C., {Weiss}, N.~O. 1986, \aap, 158, 351

\bibitem[{{Schmidt} \& {Fritz}(2004)}]{2004A&A...421..735S}
{Schmidt}, W., {Fritz}, G. 2004, \aap, 421, 735

\bibitem[{{Sch{\"u}ssler} \& {V{\"o}gler}(2006)}]{2006ApJ...641L..73S}
{Sch{\"u}ssler}, M., {V{\"o}gler}, A. 2006, \apjl, 641, L73

\bibitem[{{Solanki}(2003)}]{2003A&ARv..11..153S}
{Solanki}, S.~K. 2003, \aapr, 11, 153

\bibitem[{{Solanki} \& {Montavon}(1993)}]{1993A&A...275..283S}
{Solanki}, S.~K., {Montavon}, C.~A.~P. 1993, \aap, 275, 283

\bibitem[{{Spruit} \& {Scharmer}(2006)}]{2006A&A...447..343S}
{Spruit}, H.~C., {Scharmer}, G.~B. 2006, \aap, 447, 343

\bibitem[{{Stein} \& {Nordlund}(1998)}]{1998ApJ...499..914S}
{Stein}, R.~F., {Nordlund}, A. 1998, \apj, 499, 914

\bibitem[{{Thomas}(2005)}]{2005A&A...440L..29T}
{Thomas}, J.~H. 2005, \aap, 440, L29

\bibitem[{{Thomas} \& {Montesinos}(1990)}]{1990ApJ...359..550T}
{Thomas}, J.~H., {Montesinos}, B. 1990, \apj, 359, 550

\bibitem[{{Thomas} \& {Montesinos}(1991)}]{1991ApJ...375..404T}
{Thomas}, J.~H., {Montesinos}, B. 1991, \apj, 375, 404

\bibitem[{{Thomas} \& {Weiss}(2004)}]{2004ARA&A..42..517T}
{Thomas}, J.~H., {Weiss}, N.~O. 2004, \araa, 42, 517

\bibitem[{{Thomas} {et~al.}(2002){Thomas}, {Weiss}, {Tobias}, \&
  {Brummell}}]{2002Natur.420..390T}
{Thomas}, J.~H., {Weiss}, N.~O., {Tobias}, S.~M., {Brummell}, N.~H. 2002, \nat,
  420, 390

\bibitem[{{Thomas} {et~al.}(2006){Thomas}, {Weiss}, {Tobias}, \&
  {Brummell}}]{2006A&A...452.1089T}
{Thomas}, J.~H., {Weiss}, N.~O., {Tobias}, S.~M., {Brummell}, N.~H. 2006, \aap,
  452, 1089

\bibitem[{{Tildesley} \& {Weiss}(2004)}]{2004MNRAS.350..657T}
{Tildesley}, M.~J., {Weiss}, N.~O. 2004, \mnras, 350, 657

\bibitem[{{Weiss} {et~al.}(2004){Weiss}, {Thomas}, {Brummell}, \&
  {Tobias}}]{2004ApJ...600.1073W}
{Weiss}, N.~O., {Thomas}, J.~H., {Brummell}, N.~H., {Tobias}, S.~M. 2004, \apj,
  600, 1073

\end{thebibliography}

\end{small}

}
\end{document}